\documentclass[superscriptaddress,twocolumn,showpacs,prl,floatfix]{revtex4}

\usepackage{color}
\usepackage{tabularx}
\usepackage{epsfig}
\usepackage{amsmath}

\usepackage{graphicx}

\begin{document}

\title{Critical exponents in Ising Spin Glasses}

\author{P. H.~Lundow} 
\affiliation {Department of Theoretical Physics, Kungliga Tekniska
  h\"ogskolan, SE-106 91 Stockholm, Sweden}

\author{I. A.~Campbell}
\affiliation{Laboratoire Charles Coulomb, Universit\'e Montpellier II,
  34095 Montpellier, France}

\begin{abstract}
Extensive simulations are made of the spin glass susceptibility and
correlation length in five dimension Ising Spin Glasses (ISGs) with
Gaussian and bimodal interaction distributions.  Once the transition
temperature is accurately established using a standard criterion,
critical exponents and correction terms can be readily estimated by
extrapolating measurements made in the thermodynamic limit regime. The
data show that the critical exponents of the susceptibility $\gamma$
and of the correlation length $\nu$ depend on the form of the
interaction distribution. This observation implies that quite
generally critical exponents are not universal in ISGs.

\end{abstract}


\maketitle

\section{Introduction}

The universality of critical exponents is an important and remarkably
elegant property of standard second order transitions, which has been
explored in great detail through the Renormalization Group Theory.
The universality hypothesis states that for all systems within a
universality class the critical exponents are strictly identical and
do not depend on the microscopic parameters of the model. However,
universality is not strictly universal; there are known \lq\lq
eccentric\rq\rq models which violate the universality rule in the
sense that their critical exponents vary continuously as a function of
a control variable. The most famous example is the eight vertex model
solved exactly by Baxter \cite{baxter:71}; there are other scattered
cases.

For Ising Spin Glasses (ISGs), the form of the interaction
distribution is a microscopic control parameter. It has been assumed
that the members of the ISG family of transitions obey standard
universality rules, following the generally accepted statement that
\lq\lq Empirically, one finds that all systems in nature belong to one of a
comparatively small number of universality classes\rq\rq \cite{stanley:99}.


ISG transition simulations are much more demanding numerically than
are those on, say, pure ferromagnet transitions with no interaction
disorder. The traditional approach in ISGs has been to study the
temperature and size dependence of observables in the near-transition
region and to estimate the critical temperature and exponents through
finite size scaling relations after taking means over large numbers of
samples. Finite size corrections to scaling should be allowed for
explicitly which can be delicate. Usually it has been concluded that
the numerical data are compatible with universality
\cite{katzgraber:06,hasenbusch:08,jorg:08} even though the estimates
of the critical exponents have varied considerably from one
publication to the next (see Ref.~\cite{katzgraber:06} for a
tabulation of historic estimates).

We have estimated the critical exponents in two ISGs in dimension $5$
using a strategy complementary to the standard finite size scaling
method. First we use the Binder cumulant to estimate the critical
temperature $\beta_c$ reliably and with precision through finite size
scaling \cite{notebeta}. Then using the scaling variable and scaling
expressions appropriate for ISGs \cite{daboul:04,campbell:06} we
estimate the temperature dependence of the thermodynamic limit (ThL)
ISG susceptibility $\chi(\beta,\infty)$ and second moment correlation
length $\xi(\beta,\infty)$ over the entire paramagnetic temperature
range from $\beta=0$ to criticality. From these data we estimate the
critical exponents and the leading Wegner correction terms
\cite{wegner:72}. The numerical data show conclusively that for the
ISGs in dimension $5$ critical exponents do depend on the form of the
interaction distribution. It is relevant that it has been shown
experimentally that in Heisenberg spin glasses the critical exponents
depend on the strength of the Dzyaloshinski-Moriya interaction
\cite{campbell:10}.

The Hamiltonian is as usual
\begin{equation}
  \mathcal{H}= - \sum_{ij}J_{ij}S_{i}S_{j}
  \label{ham}
\end{equation}
with the near neighbor symmetric bimodal ($\pm J$) or Gaussian
distributions normalized to $\langle J_{ij}^2\rangle=1$. The Ising
spins live on simple [hyper]cubic lattices with periodic boundary
conditions.

A natural scaling variable for ISGs with symmetric interaction
distributions is $\tau = 1-(\beta/\beta_{c})^2$
\cite{singh:86,daboul:04}; then the standard ThL ISG susceptibility
including the leading Wegner correction term \cite{wegner:72} is
\begin{equation}
  \chi(\beta)= C_{\chi}\tau^{-\gamma}\left(1+a_{\chi}\tau^{\theta}\right)
  \label{wegchi}
\end{equation}
where $\gamma$ is the critical exponent and $\theta$ the Wegner
correction exponent, both of which are characteristic of a university
class.  As $\chi(\beta=0)=1$, $C_{\chi} = 1/(1+a_{\chi})$.  Following
a protocol well-established for the ferromagnetic case
\cite{kouvel:64,butera:02} one can define a temperature dependent
effective exponent $\gamma(\beta)=
-\partial\ln\chi(\beta)/\partial\ln\tau$ with $\gamma(\beta)$
tending to the critical $\gamma$ as $\tau \to 0$. A useful exact
infinite temperature limit rule from High Temperature Series
Expansions (HTSE) for ISGs on simple [hyper]cubic lattices in
dimension $d$ \cite{daboul:04} is $\gamma(\beta=0)=2d\beta_c^2$.
Samples of size $L$ are in the ThL regime as long as the condition $L
> \xi(\beta)$ is satisfied.  The exact ThL $\chi(\beta,\infty)$ can be
calculated for bimodal and Gaussian ISGs in any dimension using the
high temperature series terms tabulated by Daboul {\it et al.}
\cite{daboul:04} over a range of $\beta$ limited by the number of terms
($15$ for bimodal interactions and $13$ for Gaussian) whose values
have been explicitly evaluated.

The analogous natural scaling expression for the ISG second moment
correlation length $\xi(\beta)$ is \cite{campbell:06}
\begin{equation}
  \xi(\beta)/\beta = C_{\xi}\tau^{-\nu}\left(1+a_{\xi}\tau^{\theta}\right)
  \label{wegxi}
\end{equation}
or, alternatively, define $\nu(\beta) =
-\partial\ln(\xi(\beta)/\beta)/\partial\ln\tau$. The reason for the
factor $1/\beta$ is spelt out in Ref.~\cite{campbell:06}. The
$\beta=0$ limit in ISGs in simple [hyper]cubic lattices of dimension
$d$ is $\nu(\beta=0)= (d-K/3)\beta_c^2$ where $K$ is the kurtosis of
the interaction distribution.

When samples of finite size $L$ are in the ThL regime,
$\chi(L,\beta)$, $\xi(L,\beta)$ and other observables are independent
of $L$.  Working in the ThL has a number of advantages: the
temperatures studied are higher than the critical temperature so
equilibration is facilitated, the sample to sample variations are
automatically much weaker than at criticality, and there are no finite
size scaling corrections to take into account.
It can be noted that the particular critical exponent $\eta$ can be
estimated without needing $\beta_c$ as an input parameter
\cite{campbell:06}.
Otherwise for the temperature dependent effective exponents
$\gamma(\tau,L)$ and $\nu(\tau,L)$ it is important to already have an
accurate and reliable estimate of $\beta_c$ from finite size critical
data such as the familiar Binder cumulant or correlation length ratio,
or link overlap criteria \cite{lundow:13,lundow:13a}.
For the present analysis we have used the critical behavior of the
Binder cumulant $g(\beta,L)$ for estimating $\beta_c$ as among the
dimensionless variables it showed the small finite size corrections
for the ISGs studied. Results from the correlation length ratio
$\xi(\beta,L)/L$ and link overlap parameters were fully consistent
with the Binder estimate.
The limit of the ThL regime for each $L$ can be identified by
inspection; with $\beta_c$ fixed, the envelope curve for the whole set
of the ThL regime $\gamma(\tau)$ data points can be extrapolated to
$\tau=0$ to obtain an estimate of each of the critical exponents.

A particularly useful method for extending the susceptibility data to
criticality is to plot $y =
\partial\beta^2/\partial\ln\chi(\beta)$ against $x =
\beta^2$. If correction terms beyond the leading Wegner term can be
considered negligible there is an exact expression for the ThL regime:
\begin{equation}
  \frac{\partial\beta^2}{\ln\chi(\beta)} = \frac{\beta_c^2\tau(1+a_{\chi}\tau^{\theta})}{\gamma + (\gamma-\theta) a_{\chi} \tau^{\theta}}
  \label{dbsqdlns}
\end{equation}
The critical intercept $y=0$ occurs when $x=\beta_{c}^2$, and the
initial slope starting at the intercept is $\partial y/\partial x
=-1/\gamma$.  If ThL $\chi(\beta,L)$ data to sufficiently large $L$
are available and if the higher order Wegner correction terms are
indeed negligible (this should generally be the case except in the
region of very small $\beta$) then the four parameters $\beta_{c}^2,
\gamma, \theta$, and $a_{\chi}$ can in principle all be estimated from
a single fit to this plot of $\chi(\beta,L)$ data. From the generic
form of the HTSE the high temperature $x=0$ intercept is $y=1/2d$ for
an ISG in dimension $d$, whatever the interaction distribution and
whatever $\beta_c$. This reduces the number of free parameters to
three, as the condition $a_{\chi}\theta/(a_{\chi}+1)=
\gamma-2d\beta_{c}^2$ follows. In addition, if $\beta_c^2$ is already
accurately known from independent observations such as finite size
scaling, then the precision on the estimates of the other parameters
is obviously greatly improved as the fit reduces to a two free
parameter interpolation.

There is an analogous expression for $\xi(\beta)/\beta$:
\begin{equation}
  \frac{\partial\beta^2}{\ln(\xi(\beta)/\beta)} = \frac{\beta_c^2\tau(1+a_{\xi}\tau^{\theta})}{\nu + (\nu-\theta) a_{\xi} \tau^{\theta}}
  \label{dbsqdlnTxi}
\end{equation}
with the same $\beta_c^2$ and $\theta$ as for $\chi(\beta)$. The $y=0$
intercept is again $x=\beta_c^2$, with an initial slope at the
intercept equal to $\partial y/\partial x=-1/\nu$. The $x=0$
intercept is $y = 3/(3d-K)$, where again $d$ is the dimension and $K$
the interaction distribution kurtosis. Then, by analogy with
Eq.~\eqref{dbsqdlns}, the condition $a_{\xi}\theta/(a_{\xi}+1)=
\nu-((3d-K)/3)\beta_{c}^2$ holds, so $\nu$ is the only remaining free
parameter in the Eq.~\eqref{dbsqdlnTxi} fit.

The simulations were carried out using exchange Monte Carlo on $256$
samples at each size. Error bars on the finite difference derivatives
in Eqs.~\eqref{dbsqdlns} and ~\eqref{dbsqdlnTxi} are from the
bootstrap method, though it is clear that for a finite difference
derivative such estimates are not very meaningful.

\begin{figure}
  \includegraphics[width=3.in]{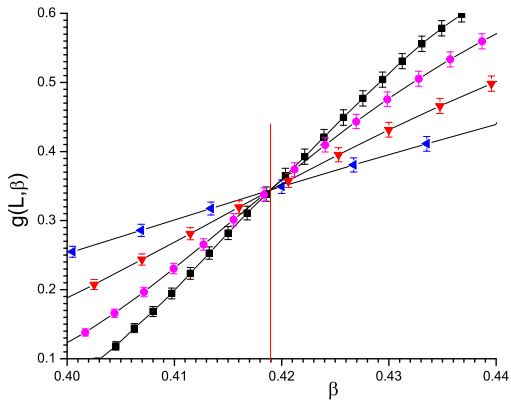}
  \caption{(Color online) The Binder cumulant $g(\beta,L)$ for even
    $L$ $5$d Gaussian interaction samples; symbol coding: black
    squares $L=10$, pink circles $L=8$, red inverted triangles $L=6$,
    blue left triangles $L=4$. The vertical red line corresponds to
    $\beta_{c}=0.419$.}\protect\label{fig:1}
\end{figure}

\begin{figure}
  \includegraphics[width=3.in]{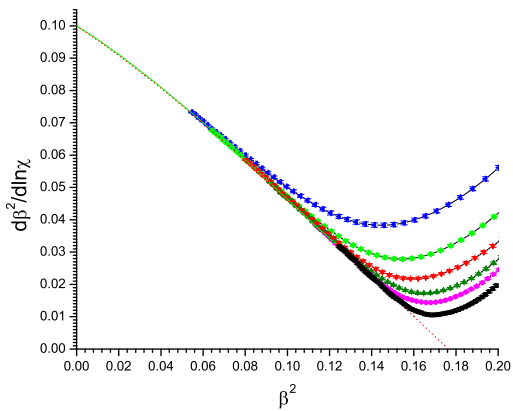}
  \caption{(Color online)
    $\partial\beta^2/\partial\ln\chi(\beta)$ for $5$d Gaussian
    interaction samples. Symbol coding as in Fig.~\ref{fig:1} plus olive
    triangle $L=7$, green diamond $L=5$. The full green curve
    calculated directly from HTSE continues to represent the ThL
    $\chi(\beta)$ up to $\beta^2 \sim 0.13$. Dashed red curve: fit
    Eq.~\eqref{dbsqdlns}. The overall ThL envelope can be seen by
    inspection.}\protect\label{fig:2}
\end{figure}

For the $5$d Gaussian ISG the HTSE critical temperature and exponent
estimates are \cite{daboul:04} $\beta_{c} = 0.4207(35)$ and $\gamma =
1.75(15)$. From the intersections of the present $g(\beta,L)$ curves
$\beta_c = 0.419(1)$, see Fig.~\ref{fig:1}. No finite size correction
for the Binder cumulant is visible, so the $\beta_c$ estimate is
particularly reliable. There is full agreement between the $\beta_{c}$
estimate from $g(\beta,L)$ and the HTSE central value, with the former
being considerably more accurate. With $\beta_c^2$ fixed at
$0.419^2=0.1755$, the optimal interpolation fit to the HTSE and
simulation ISG susceptibility data using Eq.~\eqref{dbsqdlns},
Fig.~\ref{fig:2}, is with parameters $\gamma=1.62(3), \theta =
3.0(5)$, and $a_{\chi} = -0.0445(50)$. These values are the best fit
estimates and the error bars allow for the residual uncertainty in
$\beta_c$. The final tabulated HTSE $\gamma$ estimate in
\cite{daboul:04} appears to be in only marginal agreement with the
present estimate. However, it can be noted that each individual HTSE
Dlog Pad\'{e} $\beta_{c}^2$ estimate is accompanied by a $\gamma$
estimate in almost perfect one-to-one correspondence, see Fig.~7 of
Ref.~\cite{daboul:04}. Reading off this figure, if $\beta_{c}^{2} =
0.1755$, then $\gamma \sim 1.61$. Hence there is excellent agreement
between the present $\gamma$ estimate and the HTSE Dlog Pad\'{e}
estimates. The present high $\theta$ and low $a_{\chi}$ estimates show
that Wegner correction is weak and the residual leading correction
term is of high order. This may explain why estimates from the M1
and M2 HTSE protocols \cite{daboul:04} are different from the Dlog
Pad\'{e} estimates in this particular case.

\begin{figure}
  \includegraphics[width=3.in]{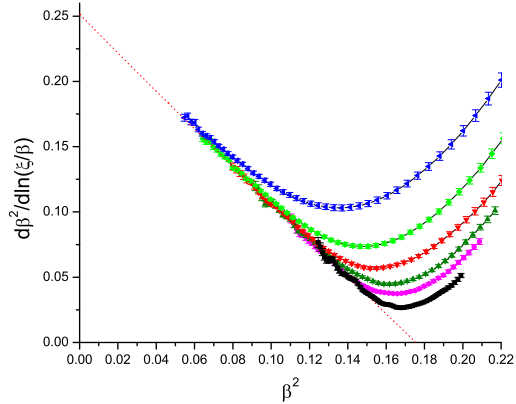}
  \caption{(Color online)
    $\partial\beta^2/\partial\ln(\xi(\beta)/\beta)$ for $5$d
    Gaussian interaction samples. Symbol coding as in
    Fig.~\ref{fig:2}.}\protect\label{fig:3}
\end{figure}

The correlation length $\xi(\beta,L)$ simulation data were analysed
following just the same procedure using Eq.~\eqref{dbsqdlnTxi}, with
$\beta_{c}^2$ and $\theta$ held fixed at the same values as estimated
above. Unfortunately there are no HTSE results available except for
the $\beta^2 = 0$ limit point. The $\xi(\beta,L)$ simulation data are
intrinsically more noisy than the $\chi(\beta,L)$ data.  The optimal
fit to the $\partial\beta^2/\partial\ln(T\xi(\beta))$ data plot
for the Gaussian interactions, Fig.~\ref{fig:3}, with the same
$\beta_{c}^2$ and $\theta$ gave the estimate $\nu = 0.71(2)$ and
$a_{\xi} = 0.004(2)$ (or $C_{\xi} \sim 1.00$). The Wegner correction
term is tiny. From the general scaling rule $\gamma = (2-\eta)\nu$, we
can estimate $\eta = -0.28(4)$.

\begin{figure}
  \includegraphics[width=3.in]{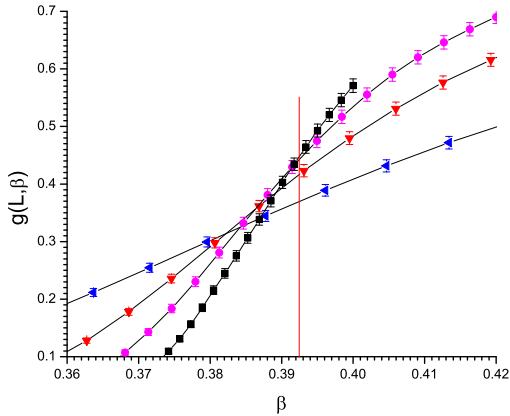}
  \caption{(Color online) The Binder cumulant $g(\beta,L)$ for $5$d
    bimodal interaction samples with color coding as in
    Fig~\ref{fig:1}. The vertical red line corresponds to $\beta_{c}=
    0.3925$.}\protect\label{fig:4}
\end{figure}

For the $5$d bimodal ISG the HTSE critical temperature estimate
\cite{daboul:04} is $\beta_{c} = 0.3925(40)$. From the intersections
of the $g(\beta,L)$ curves $\beta_c = 0.3925(10)$,
Fig.~\ref{fig:4}. The finite size $g(\beta,L)$ corrections are
stronger than in the Gaussian case. There is full agreement between
the Binder cumulant $\beta_c$ estimate and the central value of the
HTSE $\beta_c$ estimate, with the former error bars being considerably
smaller than the HTSE error bars. With $\beta_c^2$ fixed at
$0.3925^2=0.1540$, the optimal interpolation fit to the HTSE and
simulation ISG susceptibility data using Eq.~\eqref{dbsqdlns},
Fig.~\ref{fig:4}, is with fit parameters $\gamma=1.99(4), \theta =
0.88(5)$, and $a_{\chi} = 0.89$ (so $C_{\chi}=0.53$). These values are
the best fit estimates and the error bars allow for the residual
uncertainty in $\beta_c$. The agreement with the central HTSE estimate
$\gamma =1.95(15)$ is excellent but the errors on the present value
are much smaller because the estimate is based on information from
both HTSE and from simulations. The same data can be plotted as
$\gamma(\tau)$ against $\tau$ up to $\tau=0$ where a consistent
estimate of $\gamma$ is obtained, or as $\chi(\tau)\tau^{\gamma}$
against $\tau^{\theta}$, where the ThL regime data indeed fall on a
straight line with intercept $C_{\chi}=1/(1+a_{\chi})$ confirming that
the higher order Wegner corrections are negligible. It can be noted
that the HTSE analysis \cite{daboul:04} provided only a rough estimate
$\theta \sim 1.0$; no indication of the sign or value of the important
correction term strength parameter $a_{\chi}$ was given.

\begin{figure}
  \includegraphics[width=3.in]{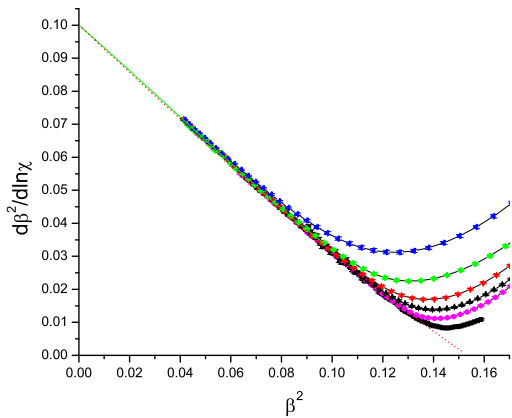}
  \caption{(Color online)
    $\partial\beta^2/\partial\ln\chi(\beta)$ for $5$d bimodal
    interaction samples. Symbol coding as in
    Fig.~\ref{fig:2}. Continuous green curve calculated from the exact
    terms in the HTSE ISG susceptibility tabulation \cite{daboul:04}
    (continues up to $\beta^2 \sim 0.10$). Dashed red curve : fit
    Eq.~\eqref{dbsqdlns}.}\protect\label{fig:5}
\end{figure}

The correlation length $\xi(\beta,L)$ simulation data were analysed
using Eq.~\eqref{dbsqdlnTxi}. The optimal fit was with $\nu = 0.86(2)$
and $a_{\xi} =0.19$ (or $C_{\xi} = 0.84$).  The analogous alternative
plots were made for $\xi(\beta,L)$ as for $\chi(\beta,L)$, and again
full consistency was observed.  The estimate $\eta= -0.32(4)$ follows
from the scaling rule $\gamma= \nu(2-\eta)$.



\section{Conclusions}
In conclusion, numerical information on finite size scaling
observables, on the ISG susceptibility and on the correlation length
from simulations has been combined with information from the exact 15
(bimodal) or 13 (Gaussian) term HTSE susceptibility tables
\cite{daboul:04} to obtain high precision empirical estimates of the
critical temperatures $\beta_c$, the critical exponents $\gamma$ and
$\nu$, and the parameters of the leading Wegner correction terms, for
the bimodal and Gaussian ISGs in dimension $5$. The $\beta_c$ values
are in full agreement with, but are considerably more precise than,
estimates from HTSE alone \cite{daboul:04}. As a result and because of
the use of a novel analysis protocol for the ThL data, the precision
on the $\gamma$ estimates is improved by a factor of the order of $5$
as compared with the estimates obtained in Ref.~\cite{daboul:04}. The
present $\nu$ estimates are of similar quality to those for $\gamma$;
there are no published $\nu$ values in dimension $5$ to compare with.

The accurate estimates of $\gamma$ and $\nu$ show that the bimodal and
Gaussian ISGs in $5$d have different critical exponents. Results in
dimension $4$ \cite{lundow:13a} and a reanalysis of data in dimension
$3$ taking special care concerning the estimates of the critical
temperatures \cite{lundow} confirm this conclusion. These results
clearly imply that in the entire family of ISGs the critical exponents
are dependent on the form of the interaction distribution, a
\lq\lq microscopic\rq\rq parameter. Other model parameters, such as a bias in
the interaction distribution, could be explored. It would obviously be
of fundamental interest to understand the basic origin of this lack of
universality at ISG transitions.

\section{Acknowledgements} 
We are very grateful to K. Hukushima for comments and communication of
unpublished data. We thank Amnon Aharony for constructive
criticism. The computations were performed on resources provided by
the Swedish National Infrastructure for Computing (SNIC) at the High
Performance Computing Center North (HPC2N).


\begin{thebibliography}{20}
\bibitem{notebeta} We will use inverse temperatures $\beta = 1/T$ throughout.
\end{thebibliography}
\end{document}